\documentclass[preprint,12pt]{article}
 
\usepackage{amssymb}
\usepackage{caption}
\usepackage{breqn}
\usepackage{graphicx}
\usepackage{amsmath} 
\usepackage{cite}

\usepackage{xparse}
\ExplSyntaxOn
\NewDocumentCommand{\mref}{m}{\quinn_mref:n {#1}}
\seq_new:N \l_quinn_mref_seq
\cs_new:Npn \quinn_mref:n #1
 {
  \seq_set_split:Nnn \l_quinn_mref_seq { , } { #1 }
  \seq_pop_right:NN \l_quinn_mref_seq \l_tmpa_tl
  ( % print the left parenthesis
  \seq_map_inline:Nn \l_quinn_mref_seq
    { \ref{##1},\nobreakspace } % print the first references
  \exp_args:NV \ref \l_tmpa_tl % print the last or only one
  ) % print the right parenthesis
 }
\ExplSyntaxOff

\begin{document}

\begin{center}

\textbf{Universal transcendentality  limit of BFKL eigenvalue}

\small
Mohammad Joubat$^{(a)}$   and Alex Prygarin$^{(b)}$  
\\ \nonumber
$^{(a)}$ Department of Mathematics, Ariel University, Ariel 40700, Israel\\ \nonumber
$^{(b)}$ Department of Physics, Ariel University, Ariel 40700, Israel 
\end{center}
\normalsize

\normalsize

\begin{abstract}
 We  consider a special limit of the BFKL eigenvalue at $\nu \to 0$  and  odd values of the conformal spin $n$. We show that in this limit  the NLO BFKL eigenvalue  can be expressed in terms of a limited set of transcendental  constants with rational coefficients. We show that the leading transcendentality term  in this limit is  universal and does not  depend  on a  specific  value of $n$. 
 
\end{abstract}
%%%%%%%%%%%%%%%%%%%%%%%%%%%%%%%%%%%%%%%%%%%%%%%%%%%%%%%%%%%
%%%%%%%%%%%%%%%%%%%%%%%%%%%%%%%%%%%%%%%%%%%%%%%%%%%%%%%%%%%
%%%%%%%%%%%%%%%%%%%%%%%%%%%%%%%%%%%%%%%%%%%%%%%%%%%%%%%%%%%
%%%%%%%%%%%%%%%%%%%%%%%%%%%%%%%%%%%%%%%%%%%%%%%%%%%%%%%%%%%
%%%%%%%%%%%%%%%%%%%%%%%%%%%%%%%%%%%%%%%%%%%%%%%%%%%%%%%%%%%
%%%%%%%%%%%%%%%%%%%%%%%%%%%%%%%%%%%%%%%%%%%%%%%%%%%%%%%%%%%
%%%%%%%%%%%%%%%%%%%%%%%%%%%%%%%%%%%%%%%%%%%%%%%%%%%%%%%%%%%
%%%%%%%%%%%%%%%%%%%%%%%%%%%%%%%%%%%%%%%%%%%%%%%%%%%%%%%%%%%
%%%%%%%%%%%%%%%%%%%%%%%%%%%%%%%%%%%%%%%%%%%%%%%%%%%%%%%%%%%

\newpage 

\section{Introduction}

The Balitsky-Fadin-Kuraev-Lipatov~(BFKL)~\cite{BFKL1,BFKL2,BFKL3,BFKL4}  equation describes the bound state of two reggeized gluons. The full functional form of the kernel of the  BFKL equation in QCD as well as the eigenfunctions and eigenvalues that solve it are currently known at the leading order~(LO) and the next-to-leading~(NLO)~\cite{NLO} order of the perturbation theory with expansion parameter being the gauge coupling constant enhanced by the corresponding power of  the  logarithm of the center-of-mass energy.   The difference between the leading order and the next-top-leading order is that the latter is suppressed by one power of   logarithm of energy. The BFKL equation is formulated for a general color state configuration.
The most studied  color configurations   are the  color adjoint and  the color singlet states of the two reggeized gluons. The color adjoint state was extensively investigated  over past decade in the context of the helicity amplitudes in $N=4$ super Yang-Mills~(SYM) 
theory~\cite{MHV7,MHV6,MHV5,MHV4,MHV3,MHV2,MHV1,FadinAdj1,FadinAdj2,FadinAdj3,FadinAdj4,Dixon1,Dixon2,Dixon3,Bondarenko:2015tba,Bondarenko:2016tws,Prygarin:2018tng}. The color adjoint BFKL eigenvalue was calculated at  higher orders of the perturbation 
theory in the context of the helicity amplitudes~\cite{MHV1,MHV2,D1,C1}. The situation with the color singlet BFKL eigenvalue is rather different. Its full functional form is known only in the next-to-leading order in both QCD and $N=4$ SYM, while at   higher orders of the perturbation theory  only some special limits of it was calculated using Quantum Spectral Curve~\cite{gromov,GromovNonzero,Alfimov:2020obh}  
 and other techniques~\cite{velizh,huot,Caron-Huot:2015bja,Caron-Huot:2020grv,Gardi:2019pmk,Vernazza:2018gyb,Caron-Huot:2017fxr}. The BFKL eigenvalue is a function of two variables, the real continuous $\nu$, related to the anomalous dimension of twist two operators, and  the conformal spin  $n$, which takes real integer values.  

In this paper we consider the next-to-leading order color singlet BFKL eigenvalue
 in $N=4$ SYM in the limit of $\nu \to 0$  for odd values of $n$ 
  and show that it    built of a limited set of transcendental numbers with rational coefficients.
We show that the leading transcendentality term is universal in this limit and does not depend on a specific value of $n$.

 \newpage 

\section{  BFKL eigenvalue}
The  BFKL eigenvalue   theory was first  calculated by Fadin and Lipatov  using Feynman diagrams in QCD and then was 
 derived in $N=4$ SYM in accordance with "the maximal transcendentality principle" by Kotikov and Lipatov~\cite{KotikovDGLAP1,KotikovDGLAP2,maxtrans}. 
In $N=4$ SYM  the  color singlet  BFKL eigenvalue is given by  
\begin{eqnarray}\label{omega}
\omega= 4 \bar{a} \left[\chi(n , \gamma)+\bar{a} \; \delta(n, \gamma)\right]
\end{eqnarray} 
where $\gamma= \frac{1}{2}-i \nu$ and the leading order~(LO) expression is given by  
\begin{eqnarray}\label{chi}
\chi(n, \gamma)= 2 \psi(1) -\psi\left(M\right)-\psi\left(1-\bar{M}\right)
\end{eqnarray}
and the next-to-leading~(NLO) BFKL eigenvalue reads
\begin{eqnarray}\label{delta}
\delta(n, \gamma)= \phi\left(M\right)+\phi\left(1-\bar{M}\right)-\frac{\omega_0}{2 \bar{a}} \left(\rho\left(M\right)+\rho\left(1-\bar{M}\right)\right).
\end{eqnarray}
The functional dependence on the twist-$2$ operators and the conformal spin enters through the complex variables 
\begin{eqnarray}
M=\gamma+\frac{|n|}{2}, \;\;\;   \bar{M}= \gamma-\frac{|n|}{2}.
\end{eqnarray} 
Here $\psi(z)$ is the Euler $\psi$-function is given by  the logarithmic derivative 
of the Gamma function  $\psi(z)=\frac{d \ln \Gamma(z)}{dz}$, and the  coupling $\bar{a}=\frac{g^2 N_c}{16 \pi^2}$ is defined  in terms of the coupling constant $g$ in the DREG scheme.
The functions $\rho(M)$ and $\phi(M)$ are given by 
\begin{eqnarray}
\rho(M) = \beta'(M)+\frac{1}{2} \zeta(2)
\end{eqnarray}   
and 
\begin{eqnarray}\label{phi}
\phi(M)= 3 \zeta(3) +\psi''(M)-2 \Phi_2(M)+2 \beta'(M)\left( \psi(1)-\psi(M)\right),
\end{eqnarray}
where $\beta'(M)$ is the derivative of the Dirichlet beta function
\begin{eqnarray}
\beta'(M) =\frac{1}{4} \left[ \psi'\left(\frac{M+1}{2}\right)- \psi'\left(\frac{M}{2}\right) \right]=-\sum_{r=0}^{\infty} \frac{(-1)^r}{(M+r)^2}
\end{eqnarray}
and $\Phi_2 (M)$ reads
\begin{eqnarray}
\Phi_2 (M)= \sum_{k=0}^{\infty} \frac{\beta'(k+1)}{k+M}+\sum_{k=0}^{\infty} \frac{(-1)^k\psi'(k+1)}{k+M}-\sum_{k=0}^{\infty} \frac{(-1)^k\left( \psi(k+1)-\psi(1)\right)}{(k+M)^2}.
\end{eqnarray}

The expressions of $\chi(n,\gamma)$ and $\delta(n,\gamma)$ are real valued  functions of real integer $n$ and complex  $\gamma= \frac{1}{2}-i \nu$   for  real $\nu$. 

\section{Transcendental constants, transcendentality and the loop order}
In this section we give a brief review of the transcendentality concept we use in our analysis. We start with the definition of the transcendental number and then extend the notion of transcendentality to the functions that appear in the BFKL eigenvalue. 

The transcendental number is an irrational number that is not a root of any polynomial with rational coefficients. For example,  $\pi$ is transcendental number while $\sqrt{2}$ is irrational, but not transcendental because it is  a solution to the polynomial equation $x^2-2=0$. Another useful number $\ln 2$ is also a transcendental number.  In fact, most of numbers are transcendental, but we are interested in those that can appear in the BFKL eigenvalue.   Consider the Riemann zeta function $\zeta(n)$, for even values of $n$ it can be written in terms of the powers of $\pi$ and thus it gives transcendental numbers for even values of $n$. For odd values of $n$ it gives irrational numbers that are strongly suspected to be transcendental, though a rigorous mathematical proof of this is still to be found. 
Polylogarithmic functions of half integer argument $\mathtt{Li}_s\left(\frac{1}{2} \right)$ also give irrational numbers that appear in some limits of the BFKL eigenvalue. They are  usually treated as transcendental numbers despite the lack of rigorous mathematical proof that they are such. In fact, $\mathtt{Li}_s\left(\frac{1}{2} \right)$ is  reducible  for $s<  4$ and can be written in terms of  transcendental  numbers $\ln 2 $, $\pi$ and $\zeta(3)$. For the purpose of the present discussion we treat all of the above-mentioned numbers as transcendental numbers. 

Out of infinitely many transcendental numbers there are only two constants $\pi^2 $ and $\zeta(3)$  that  appear in the BFKL eigenvalue together with polygamma functions and their generalizations.  The origin of those two constants  is better understood if one writes the NLO BFKL eigenvalue in terms of  the nested harmonic sums~(see Appendix). In the limit of $\nu \to \infty$ the BFKL eigenvalue must reproduce the expansion of the cusp anomalous dimension, which is expressed in terms of $\zeta(n)$. On the other hand the nested harmonic sums also give the same set of transcendental constants at the infinite value of the argument. This restricts the possible choice of the transcendental numbers that are relevant for the BFKL eigenvalue. 

At each loop order of the BFKL eigenvalue the set of possible transcendental numbers is limited by the principle of maximal transcendentality. At the leading order the maximal transcendentality is one, at the next-to-leading order the maximal transcendentality of constants is three, at the next-to-next-to-leading order the maximal transcendentality is five etc. The transcendentality changes by units of two from order to order in the perturbation theory. For example, the set of possible transcendental numbers that can appear at the next-to-next-to-leading BFKL eigenvalue reads
\begin{eqnarray}\label{list}
\{\pi^2, \zeta(3), \mathtt{Li}_4\left( \frac{1}{2}\right), \zeta(5)\}. 
\end{eqnarray}
All of them appear in the limit of the infinite value of the argument of the nested harmonic sums. The constant $\ln 2$ is also allowed because it appears in the limit of $ \lim_{n \to \infty}S_{-1}(n)=-\ln 2$, but it is absent in the BFKL eigenvalue. This might be related to the fact that in the NLO case  as well as in the known NNLO expressions the nested harmonic sums never have index $-1$. Most probably, this fact in its turn is related to the cylindrical topology~(closed spin chain) of the singlet BFKL equation. 

In the next section we consider a new limit of $\nu \to 0$ for odd values of the conformal spin $n$ in which the NLO BFKL eigenvalue gives a constant that can be written in terms of the transcendental  numbers in eq.~(\ref{list}). 

 \section{The uniform transcendentality limit}
We consider the limit of the NLO eigenvalue $\delta(n,\gamma)$ in which  $\nu \to 0$
and the conformal spin $n$ takes special odd values  given by $n=4k+1~(k=0,1,2,3,...)$. In this limit  the variable $\gamma= \frac{1}{2}-i \nu $  equals one half and  the first fifteen terms of  $\delta(n,\gamma)$ read~
\begin{eqnarray}
&&   \delta\left( 1,\frac{1}{2}\right)= 0
\\ \nonumber
&&   \delta\left( 5,\frac{1}{2}\right)= 6+\pi ^2
\\ \nonumber
&&   \delta\left( 9,\frac{1}{2}\right)= \frac{725}{108}+\frac{25 \pi ^2}{18}
\\ \nonumber
&&   \delta\left( 13,\frac{1}{2}\right)= \frac{31213}{4500}+\frac{49 \pi ^2}{30}
\\ \nonumber
&&   \delta\left(17 ,\frac{1}{2}\right)=\frac{86873597}{12348000}+\frac{761 \pi ^2}{420}
\\ \nonumber
&&   \delta\left(21 ,\frac{1}{2}\right)= \frac{7089858149}{1000188000}+\frac{7381 \pi ^2}{3780}
\\ \nonumber
&&   \delta\left( 25 ,\frac{1}{2}\right)= \frac{9479055229969}{1331250228000}+\frac{86021 \pi ^2}{41580}
\\ \nonumber
&&   \delta\left( 29,\frac{1}{2}\right)= \frac{20886136478164093}{2924756750916000}+\frac{1171733 \pi ^2}{540540}
\\ \nonumber
&&   \delta\left( 33,\frac{1}{2}\right)= \frac{167423375594334947}{23398054007328000}+\frac{2436559 \pi ^2}{1081080}
\\ \nonumber
&&   \delta\left( 37,\frac{1}{2}\right)= \frac{274577795894479279537}{38318213112667488000}+\frac{14274301 \pi ^2}{6126120}
\\ \nonumber
&&   \delta\left( 41,\frac{1}{2}\right)= \frac{75413751118241146234843}{10512984949591452007680}+\frac{55835135 \pi ^2}{23279256}
\\ \nonumber
&&   \delta\left(45 ,\frac{1}{2}\right)= \frac{75475515884701985008123}{10512984949591452007680}+\frac{6364399 \pi ^2}{2586584}
\\ \nonumber
&&   \delta\left(  49,\frac{1}{2}\right)= \frac{918900025283615884381544531}{127911487881679196577442560}+\frac{1347822955 \pi ^2}{535422888}
\\ \nonumber
&&   \delta\left( 53,\frac{1}{2}\right)= \frac{2873035558259746167005637883471}{399723399630247489304508000000}+\frac{34395742267 \pi ^2}{13385572200}
\\ \nonumber
&&   \delta\left( 57,\frac{1}{2}\right)= \frac{232812878912645811878247805986151}{32377595370050046633665148000000}+\frac{315404588903 \pi ^2}{120470149800}.
\end{eqnarray}
A similar structure persists for higher values of the conformal spin $n$ we have considered.   One can notice that the most transcendental terms of  either $\zeta(3)$ or $\zeta(2)\ln 2 $ allowed at this perturbative order are absent.
The $\pi^2$ terms are related to $\zeta(2)=\frac{\pi^2}{6}$
 and $\zeta(-2)=-\frac{\pi^2}{12}$, which are of the lower 
transcendentality.  We have checked other values of the conformal spin in this limit and found that  the most transcendental terms of the color singlet  BFKL eigenvalue in either QCD or $N=4$ SYM are absent for     $\nu \to 0$ and  $n=4k+1,~\;(k=0,1,2,3,...)$. The  list of the expression for  $\delta(n,\gamma)$ in this limit for the first thousand values of $n=4k+1~(k=0,1,2,3,...)$ is given in the attached file.

Next we consider a similar limit of $\nu \to 0$ for $n=4k+3,~\;(k=0,1,2,3,...)$. In this case the most transcendental term is non-zero and has a universal  coefficient  for all terms we have calculated.  The  first fifteen terms of the color singlet NLO BFKL eigenvalue in the    limit of $\nu \to 0$ for $n=4k+3,~\;(k=0,1,2,3,...)$ are given by
\begin{eqnarray}
&&  \delta\left(3 ,\frac{1}{2}\right)= 4 \zeta (3)+\frac{2 \pi ^2}{3}
\\ \nonumber
&&  \delta\left(7 ,\frac{1}{2}\right)= 4 \zeta (3)+\frac{5}{3}+\frac{11 \pi ^2}{9}
\\ \nonumber
&&   \delta\left(11 ,\frac{1}{2}\right)= 4 \zeta (3)+\frac{49}{24}+\frac{137 \pi ^2}{90}
\\ \nonumber
&&   \delta\left( 15,\frac{1}{2}\right)= 4 \zeta (3)+\frac{787}{360}+\frac{121 \pi ^2}{70}
\\ \nonumber
&&   \delta\left( 19,\frac{1}{2}\right)=
4 \zeta (3)+\frac{81917}{36288}+\frac{7129 \pi ^2}{3780}
\\ \nonumber
&&    \delta\left(23 ,\frac{1}{2}\right)= 4 \zeta (3)+\frac{10424089}{4536000}+\frac{83711 \pi ^2}{41580}
\\ \nonumber
&&    \delta\left( 27,\frac{1}{2}\right)= 4 \zeta (3)+\frac{16562497}{7128000}+\frac{1145993 \pi ^2}{540540}
\\ \nonumber
\end{eqnarray}

\begin{eqnarray}
\nonumber
&&    \delta\left( 31,\frac{1}{2}\right)= 4 \zeta (3)+\frac{14879171647}{6356750400}+\frac{1195757 \pi ^2}{540540}
\\ \nonumber
&&    \delta\left( 35,\frac{1}{2}\right)= 4 \zeta (3)+\frac{119646854891}{50854003200}+\frac{42142223 \pi ^2}{18378360}
\\ \nonumber
&&    \delta\left(39 ,\frac{1}{2}\right)= 4 \zeta (3)+\frac{6124929706001}{2593554163200}+\frac{275295799 \pi ^2}{116396280}
\\ \nonumber
&&   \delta\left( 43,\frac{1}{2}\right)= 4 \zeta (3)+\frac{10609392277369}{4479775372800}+\frac{18858053 \pi ^2}{7759752}
\\ \nonumber
&&    \delta\left( 47,\frac{1}{2}\right)= 4 \zeta (3)+\frac{14152043794450139}{5962581021196800}+\frac{444316699 \pi ^2}{178474296}
\\ \nonumber
&&    \delta\left( 51,\frac{1}{2}\right)= 4 \zeta (3)+\frac{125408026665021137}{52745909033664000}+\frac{34052522467 \pi ^2}{13385572200}
\\ \nonumber
&&    \delta\left( 55 ,\frac{1}{2}\right)= 4 \zeta (3)+\frac{107963113217545418747}{45345428666201664000}+\frac{312536252003 \pi ^2}{120470149800}
\\ \nonumber
&&    \delta\left( 59,\frac{1}{2}\right)= 4 \zeta (3)+\frac{2485984260732243834301}{1042944859322638272000}+\frac{9227046511387 \pi ^2}{3493634344200}
\end{eqnarray}
One can note the most transcendental term $\zeta(3)$ has a universal coefficient of $4$ that  does not depend on a specific value of $n$. 
This pattern persists for all other values of $n=4k+3,~\;(k=0,1,2,3,...)$ we have considered. Despite the lack of a general proof of the universality of the leading transcendental term, the direct calculation shows that this coefficient is the same  for $\nu \to 0$ and $n=4k+3,~\;(k=0,1,2,3,...)$. 

In our calculations we used the nested harmonic sums analytically continued from the even positive integer values of the argument to the complex plane. The details of the calculations are given in the Appendix. 

Two other limits of $a)$ $\nu \to 0$ for $n=1$ in  which the BFKL eigenvalue vanishes,  and $b)$ the limit of  and $\nu \to \infty$   that recovers the coefficients of the cusp anomalous dimension are known for many years and  used for either defining constraints or cross checking  the  results in calculations using Quantum Spectral Curve~\cite{gromov,GromovNonzero,Alfimov:2020obh} and other novel techniques~\cite{velizh,huot,Caron-Huot:2015bja,Caron-Huot:2020grv,Gardi:2019pmk,Vernazza:2018gyb,Caron-Huot:2017fxr}. The new  universal limit of $\nu \to 0$ for odd values of $n$   supplements existing constraints in fixing free coefficient of general ansatz for higher order BFKL eigenvalues. At this point we are unable to comment on recent results for the BFKL eigenvalue derived using the Quantum Spectral Curve  
 and other techniques due to the complexity of the underlying expressions. We believe that a similar universality of the most transcendental terms in the limit $\nu \to 0$ for odd values of the conformal spin $n$     also holds   beyond NLO order of the BFKL eigenvalue in QCD and $N=4$ SYM.

A few words to be said about a similar limit of $\nu \to 0 $ for even values of the conformal spin $n$. In this limit the NLO BFKL eigenvalue will give a new set of transcendental constants~(Catalan number $C_n$, polylogs of complex argument $\mathtt{Li}_s\left(\frac{1}{2}+ \frac{i}{2} \right)$ etc.). Those constants do not appear in the functional basis of nested harmonic sums and their relations should be investigated further in more details. In particular, it is not clear  what would be the irreducible and linear independent~\footnote{By linear independence of two transcendental numbers we mean that they cannot be related to each other using only rational coefficients. } set of those constants for any given transcendentality.

\section{Conclusions}
We discussed a new limit of the BFKL eigenvalue at the leading order~(LO) and the next-to-leading~(NLO) perturbative order. We show that in the limit of $\nu \to 0 $ for odd values of the conformal spin $n$ the NLO BFKL eigenvalue is a non-zero constant that can be expressed in terms of transcendental and rational numbers.  We show that for special odd values of the conformal spin  given by  $n=4k+3,~\;(k=0,1,2,3,...)$  the NLO BFKL eigenvalue gives the maximally transcendental term 
that is equal to $4 \zeta(3)$ and   does not depend on a specific value of $n$. For another values  of odd conformal spin  $n=4k+1,~\;(k=0,1,2,3,...)$ in the limit of $\nu \to 0$ the maximally transcendental term  is absent in the   NLO BFKL eigenvalue, i.e. the  coefficient of either $\zeta(3)$ or $\pi^2 \ln 2 $ is zero and also does not depend on specific value of the conformal  spin.  This universality in  coefficients of the most transcendental terms for odd values of the conformal spin suggests that there might be some other non-trivial  limits of  the BFKL eigenvalue at higher orders of the perturbation theory.

\section{Acknowledgments}
 We are indebted to Sergey Bondarenko and Ian Balitsky for enlightening discussions and valuable comments.  
 
\section*{Appendix} 
 The next-to-leading order BFKL eigenvalue includes some complicated functions that build $\phi(M)$ in eq.~(\ref{phi}). Those functions are functions of one complex variable and we found it useful to express $\phi(M)$ in terms of the nested harmonic sums analytically continued   from even integer values of the argument to the complex plane. 

 The nested harmonic sums are defined~\cite{HS1,Verm1998uu,Blum1998if,Rem1999ew} in terms of  nested summation for $n\in \mathbb{N}$
\begin{eqnarray}\label{defS}
S_{a_1,a_2,...,a_k}(n)=  \sum_{n \geq i_1 \geq i_2 \geq ... \geq i_k \geq 1 }   \frac{\mathtt{sign}(a_1)^{i_1}}{i_1^{|a_1|}}... \frac{\mathtt{sign}(a_k)^{i_k}}{i_k^{|a_k|}}.
\end{eqnarray}
The harmonic sums  are defined for non-zero real integer values of $a_i$, that   build the alphabet of the possible negative and positive indices uniquely  labeling $S_{a_1,a_2,...,a_k}(n)$.  
It is useful to define two important characteristics for the harmonic sums. Namely, the \emph{depth} $k$ that denotes the number of the nested summations  and the \emph{weight} $w=\sum_{i=1}^{k}|a_i|$ that is related to the transcendentality of a given sum. The nested harmonic sums are defined for positive integer values of the argument and require analytic continuation to the complex plane done using their integral representation.  There are two different analytic continuations $a)$ from even integer values of the argument  and $b)$ from the odd integer values of the argument to the complex plane. The choice of the analytic continuation is a matter of convenience and we choose the continuation from the even integers denoted by $\tilde{S}^+$ consistent with recent works on the BFKL eigenvalue.  The details of two analytic continuations 
can be found in the work of Kotikov and Velizhanin~\cite{Velizhanin}.  The analytic properties of the nested harmonic sums in the context of the BFKL eigenvalues were recently studied in a series of publications~\cite{Prygarin:2018cog,Prygarin:2019ruv,Joubat:2019esj,Joubat:2020hvc,Joubat:2020vrw}. 

The most complicated part of the NLO BFKL eigenvalue is given by the function $\phi(M)$ in eq.~(\ref{phi}) 
 \begin{eqnarray}\label{phiApp}
\phi(M)= 3 \zeta(3) +\psi''(M)-2 \Phi_2(M)+2 \beta'(M)\left( \psi(1)-\psi(M)\right),
\end{eqnarray}
where  $\Phi_2 (M)$ reads
\begin{eqnarray}
\Phi_2 (M)&=& \sum_{k=0}^{\infty} \frac{\beta'(k+1)}{k+M}+\sum_{k=0}^{\infty} \frac{(-1)^k\psi'(k+1)}{k+M}
\\
&& -\sum_{k=0}^{\infty} \frac{(-1)^k\left( \psi(k+1)-\psi(1)\right)}{(k+M)^2}. \nonumber
\end{eqnarray}

The complex variable $M=\gamma+\frac{|n|}{2}$ is defined in terms $\gamma=\frac{1}{2}-i \nu$ for real continuous $\nu$ and real integer $n$. 

The main objective of the present study is to consider the NLO BFKL eigenvalue in the limit of $\nu \to 0$ and odd values of the conformal spin $n$. In this limit the variable  $M=\frac{1}{2}+\frac{|n|}{2}$ is a positive integer and its convenient to express the function $\phi(M)$ 
 in terms of the nested harmonic sums that are easy to compute. 
 
The expression of $\phi(M)$ written  in terms of the nested harmonic sums is given by
\begin{eqnarray}
\phi(M)= -4 S_{-2,1}+2 S_{-3} +\frac{\pi^2}{3} S_{1} +4 S_{-2} S_{1}+2 S_{3},
\end{eqnarray}
where for simplicity each sums $S_{a}$ denotes the analytically continued nested harmonic sum from the even integer values of the argument to the complex plane, namely $\tilde{S}^+_{-2,1}(M-1)$. The shift of the argument by unity is due to the respective definitions of the harmonic sums and the functions of $\phi(M)$. The rest of terms in the NLO eigenvalue in $\delta(n, \gamma)$ are easily calculable by Mathematica using built-in libraries of special functions.

\end{document}